\documentclass[conference]{IEEEtran}
\usepackage{graphicx}
\usepackage{multirow}
\PassOptionsToPackage{hyphens}{url}\usepackage{hyperref}
\usepackage{url}
\usepackage{subcaption}
\usepackage{calc}
\usepackage{amsthm}
\usepackage{amsmath}
\usepackage{algorithm}
\usepackage[noend]{algpseudocode}
\usepackage{algorithmicx}
\usepackage{xcolor}
\usepackage{multirow}
\usepackage{fancyhdr}
\begin{document}

\title{An NWDAF Approach to 5G Core Network Signaling Traffic: Analysis and Characterization}
%
%
%

\author{Dimitrios~Michael~Manias, Ali~Chouman, and Abdallah~Shami\\
ECE Department, Western University, London ON, Canada\\
\{dmanias3, achouman, Abdallah.Shami\}@uwo.ca}


\maketitle

\begin{abstract}
Data-driven approaches and paradigms have become promising solutions to efficient network performances through optimization. These approaches focus on state-of-the-art machine learning techniques that can address the needs of 5G networks and the networks of tomorrow, such as proactive load balancing. In contrast to model-based approaches, data-driven approaches do not need accurate models to tackle the target problem, and their associated architectures provide a flexibility of available system parameters that improve the feasibility of learning-based algorithms in mobile wireless networks. The work presented in this paper focuses on demonstrating a working system prototype of the 5G Core (5GC) network and the Network Data Analytics Function (NWDAF) used to bring the benefits of data-driven techniques to fruition. Analyses of the network-generated data explore core intra-network interactions through unsupervised learning, clustering, and evaluate these results as insights for future opportunities and works.
\end{abstract}


\section{Introduction}

The first Fifth Generation (5G) standard, specified by the Third Generation Partnership Project (3GPP) in Release 18, was finalized in July 2018, and detailed the conditions of 5G mobile wireless communications as well as non-standalone operations, standalone operations, and 5G New Radio (NR) specifications. This laid the ground for the 5G era of wireless communications, where mission-critical applications would need to be serviced by an ever-evolving network; however, 5G and Beyond networks are posed with the difficulties of satisfying user and network operator requirements in challenging scenarios. 5G mobile communications networks face a formidable challenge in servicing the massive number of mobile phones in use today and the copious amounts of data they produce. Both the volume and the variety of massive mobile data are increasing exponentially \cite{zheng2016big} and therefore, big data in mobile networks posit the need to extensively analyze them for the purpose of easing network maintenance operations and improving future network performance. A key point discussed for improving the future performance of 5G networks is data-driven approaches, which provide scalable capacity and the performance capabilities necessary for data collection, storage, and network optimization; the 3GPP have specified the Network Data Analytics Function (NWDAF) to perform such feats \cite{3gppR16}. The NWDAF is a Network Function (NF) that uses machine learning and artificial intelligence algorithms, as an engine, to drive data analytics for the past, present, and future states of a 5G Core, while supporting other 5G Control Plane NFs in their operations. Such examples include the Policy Control Function (PCF) for steering traffic policies and the Network Slice Selection Function (NSSF) for instantiating and selecting network slices according to the data input \cite{sevgican2020intelligent}.\par

In the context of data-driven network optimization, the big data in 5G mobile networks can, generally, be divided into user data and network operator data. User data is collected from the User Equipment (UE) devices and establishes their profiles and behaviours, while also tracking metrics such as their mobility, location, and communication patterns. Due to the excessive amounts of application data produced by numerous UE devices, this application-level user data forms the incentive for rapid expansion of mobile networks' performance. Network operator data is collected from interactions between the 5G Core and the Radio Access Network (RAN); it contains an abundant amount of service data regarding network performance, handover reports, faults, link utilization, resource control messages, connection statuses, received power, signal quality, \textit{etc.} Advanced data analytics techniques in the data-driven approach to network optimization use a plethora of optimization algorithms and adaptive algorithms to manifest optimized decisions for the future of the 5G network \cite{ma2020survey}.\par

To the best of our knowledge, there are no functional 5G core and NWDAF experimental prototypes. Up until this point, previous works have relied on using 4G cellular and internet traffic to extrapolate relationships for 5G core networks \cite{alawe2018smart}. With the development of a functional prototype and the collection of a dataset specific to the 5G core, future research in the field can use realistic data for various network management decisions such as NF placement, scaling, \textit{etc.}\par
\medskip
The contributions of this paper are summarized as follows:
\begin{itemize}
\item The development of a functional NWDAF prototype.
\item The analysis of network-generated data to explore core NF-NF interactions. 
\item The use of unsupervised learning to group core NF-NF interactions.
\item A discussion on the application of NWDAF insights for intelligent network MANO.
\end{itemize}

The remainder of this paper is structured as follows. Section II outlines the 5G Core prototype and system environment. Section III outlines the methodology followed. Section IV presents and analyzes the obtained results. Section V discusses insights and applications of data-driven network intelligence. Finally, Section VI concludes the paper and discusses future work.

\section{5G Core Prototype}

\begin{figure}[!htbp]
\centerline{\includegraphics[width=0.95\columnwidth]{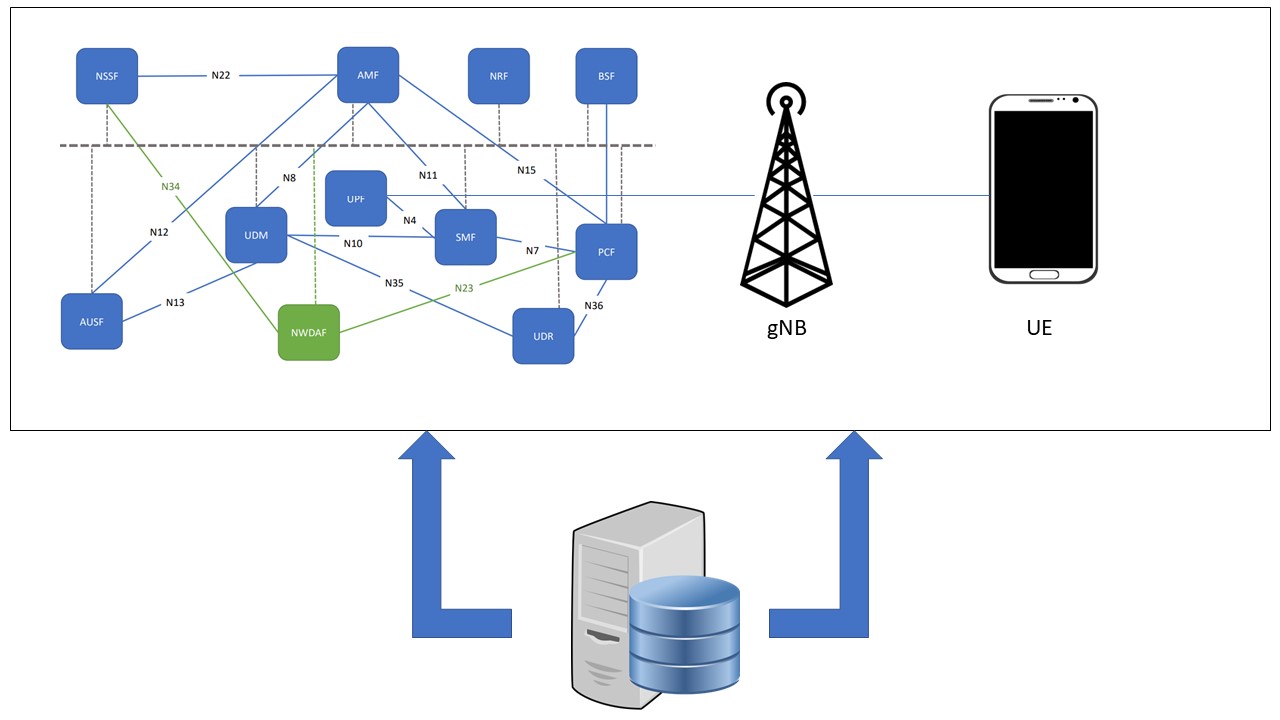}}
\caption{System Environment of 5G Core Prototype}
\label{prototype}
\end{figure}

To realize the complete emulation of the 5GC standalone architecture (SA), a single server runs a virtual management service, hosting individual Linux (Ubuntu) virtual machines (VMs) for each NF in the 5GC or entity that interacts with the 5GC. The 5GC prototype uses Open5GS, which is a C-language open-source project of the 5GC and Evolved Packet Core (EPC). Open5GS is Release-16 compliant in accordance with the 3GPP release specifications \cite{Open5GS}. Figure \ref{prototype} demonstrates the system environment as a single server configuration with multiple hosted VMs on a private network; it also shows all 5GC NFs that Open5GS provides functionalities for. The core NFs provided include the following: Network Repository Function (NRF), Access and Mobility Function (AMF), Session Management Function (SMF), Authentication Server Function (AUSF), Unified Data Management (UDM), Unified Data Repository (UDR), Policy Control Function (PCF), Network Slice Selection Function (NSSF), Binding Support Function (BSF), and the User Plane Function (UPF). In order to setup a RAN and multiple UE devices to interact with the fully operational 5GC, UERANSIM is used, which is an open-source 5G SA UE and RAN C++ implementation \cite{aligungr}. A single gNB connects to the 5GC using the aforementioned implementation and a single host uses multiple network interfaces to emulate different UE devices interacting with the RAN and the User Plane Function. All NFs are executed as Linux services or executables in each VM.\par

The implementation of the NWDAF in the system prototype uses network monitoring and data collection techniques to synthesize all service operations of the NWDAF, according to its 3GPP specification. Hyper-V, as a Type 1 Hypervisor for the server's Virtual Machine Management Service, allows port mirroring to function across all VMs designated as a source for network traffic, which duplicates all network traffic contained in the private network and forwards them to a single host as the destination, acting as the central point for NWDAF analytics and operations. Apache Kafka is used to monitor and pipeline the captured network data, and stream the data to a MongoDB instance, using Kafka Connect, to perform historical aggregate data analytics, as well as current state monitoring for future policy decision-making in the 5GC. Open5GS is operated without Network Address Translation (NAT) in the private network, and port forwarding rules are configured so that UE device traffic can be encapsulated in the GPRS Tunneling Protocol (GTP) when communicating with the UPF. The UPF can thereby route Internet connectivity to UE devices for the purpose of running sample applications on UE devices.\par

The most important advantage of using data collection and siphoning techniques through Apache Kafka is that network monitoring, through the NWDAF in the system prototype, can automatically update current packet capture records as the 5GC is operational. When multiple UE devices connect to sample applications through the 5GC, newly generated packets are processed and transformed into schema-validated NWDAF events, which can quickly comprise a dataset for any machine learning model or other algorithm. As the system prototype is configured in the aforementioned fashion, closed-loop automation is a capability, in its entirety, to maximize the potential of the NWDAF and its impact on maintenance and operation-specific decisions in the 5GC. \par

\section{Methodology}
The following section outlines the methodology followed for this work, including the data collection and analysis process as well as the interaction characterization through clustering. Regarding the data collection process, the aforementioned system environment was run for 138 minutes, and a total of 171,821 packets were captured \cite{chouman2022}. \footnote[1]{\url{https://github.com/Western-OC2-Lab/5G-Core-Networks-Datasets}} As the objective of this work is to explore, analyze and characterize the interaction between core NFs based on data collected by the NWDAF, the packets were filtered to include only those with source and destination belonging to one of the core NFs. \par
The first stage of the data analysis categorized each of the collected packets based on their NF source and destination and summary statistics, including the average packet length, the maximum packet length, the standard deviation of the packet lengths, as well as the total number of packets for each NF-NF interaction. These statistics were selected as they provide a good overview of the data and bring attention to various trends and insights related to the interactions. Specifically, as explained in the result section, certain NFs interactions are explored further, such as the NRF’s periodic interaction with the other core NFs. This further exploration enabled a greater depth of knowledge of the network as the observed trends were directly tied to core NF functionalities and processes. \par
The final objective of this work is the characterization of core NF interactions. This was accomplished by leveraging unsupervised learning, and the k-Means clustering algorithm with a variable number of clusters set a priori. This algorithm was selected for its low complexity, scalability, convergence guarantees, as well as the ability to set the number of clusters manually for analysis purposes \cite{morissette2013k}. Additionally, since dealing with a small number of clusters with a small and predefined number of dimensions, the clustering was performed multiple times to mitigate one of the main limitations of k-Means, being the high correlation between performance and initial cluster value selection \cite{celebi2013comparative}. The statistics extracted throughout the data analysis process, including the average, maximum, and standard deviation of packet length as well as the total number of packets sent, were used as features for the clustering. These features were each scaled to ensure the magnitude of the features is comparable without bias. The number of clusters, k, explored ranged from two to seven. The results of this clustering grouped similar NF-NF interactions with the intention of using these groupings to assist the Management and Orchestration (MANO) decision-making process.\par

\section{Results}
The following section presents the various results of this work, including control packet statistics, NFR interactions, as well as the NF-NF characterization through clustering. 
\subsection{Control Packet Statistics}
\begin{figure}[!htbp]
\centerline{\includegraphics[width=0.9\columnwidth]{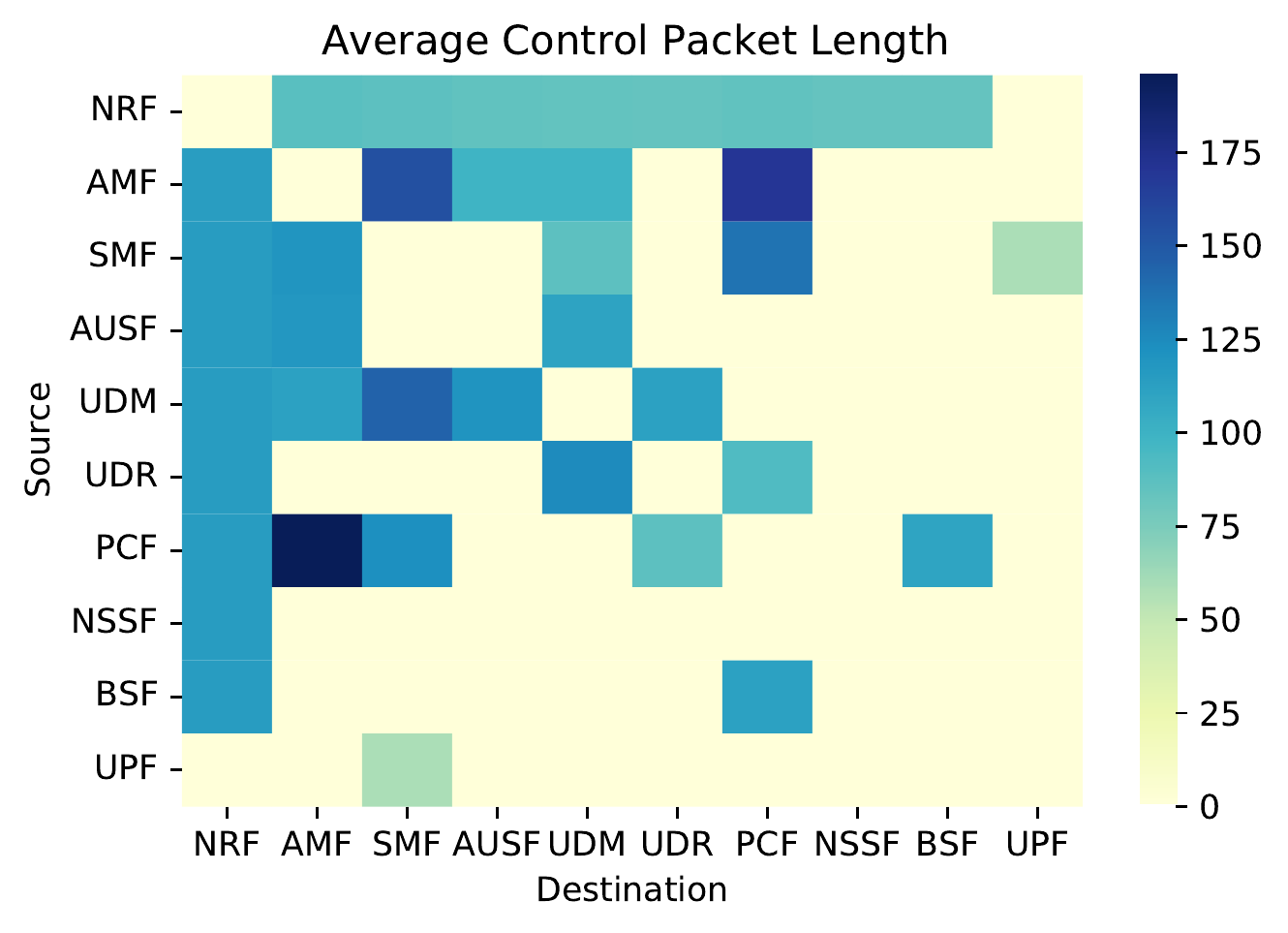}}
\caption{Average packet length per NF-NF interaction}
\label{average}
\end{figure}

\begin{figure}[!htbp]
\centerline{\includegraphics[width=0.9\columnwidth]{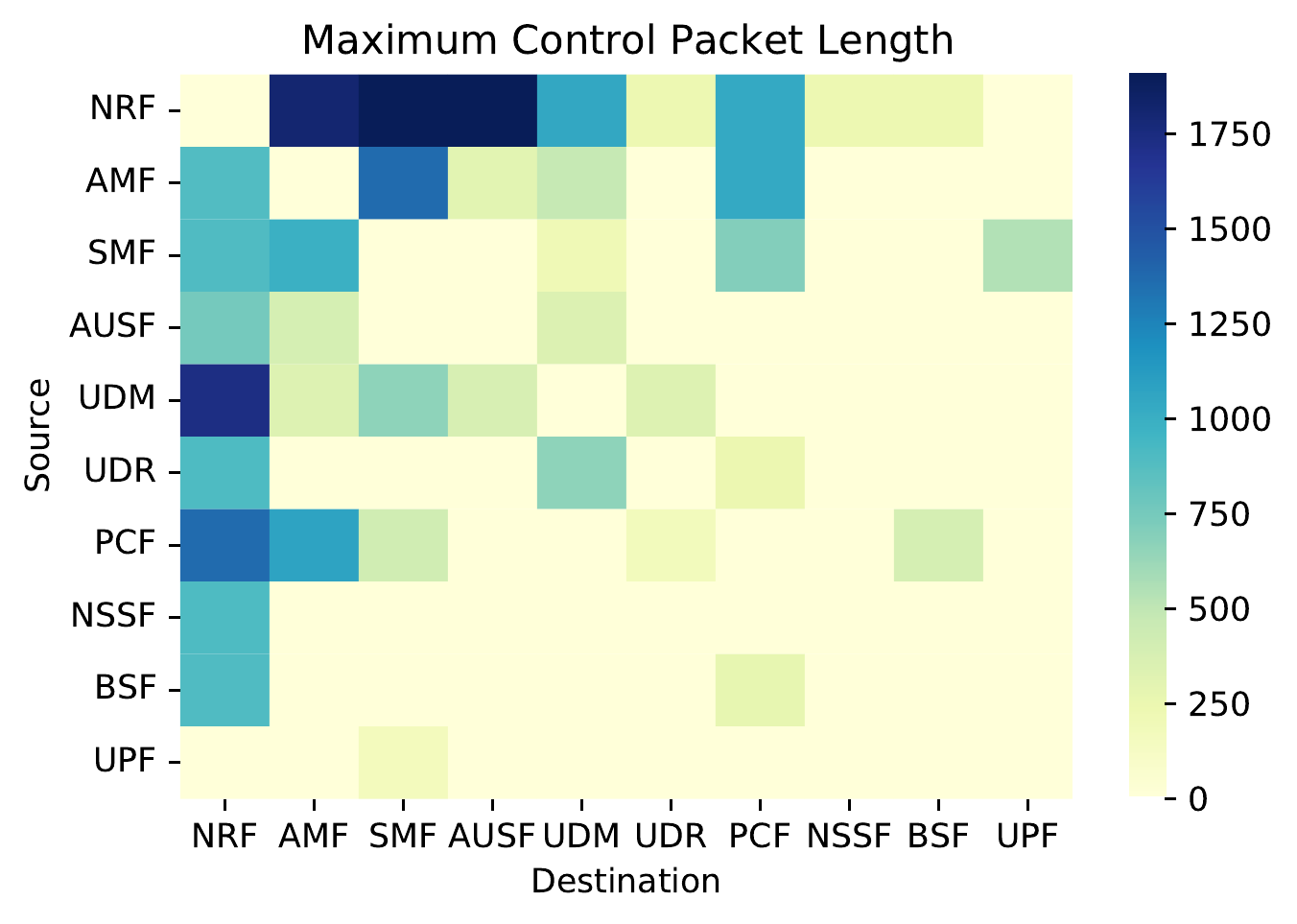}}
\caption{Maximum packet length per NF-NF interaction}
\label{max}
\end{figure}

\begin{figure}[!htbp]
\centerline{\includegraphics[width=0.9\columnwidth]{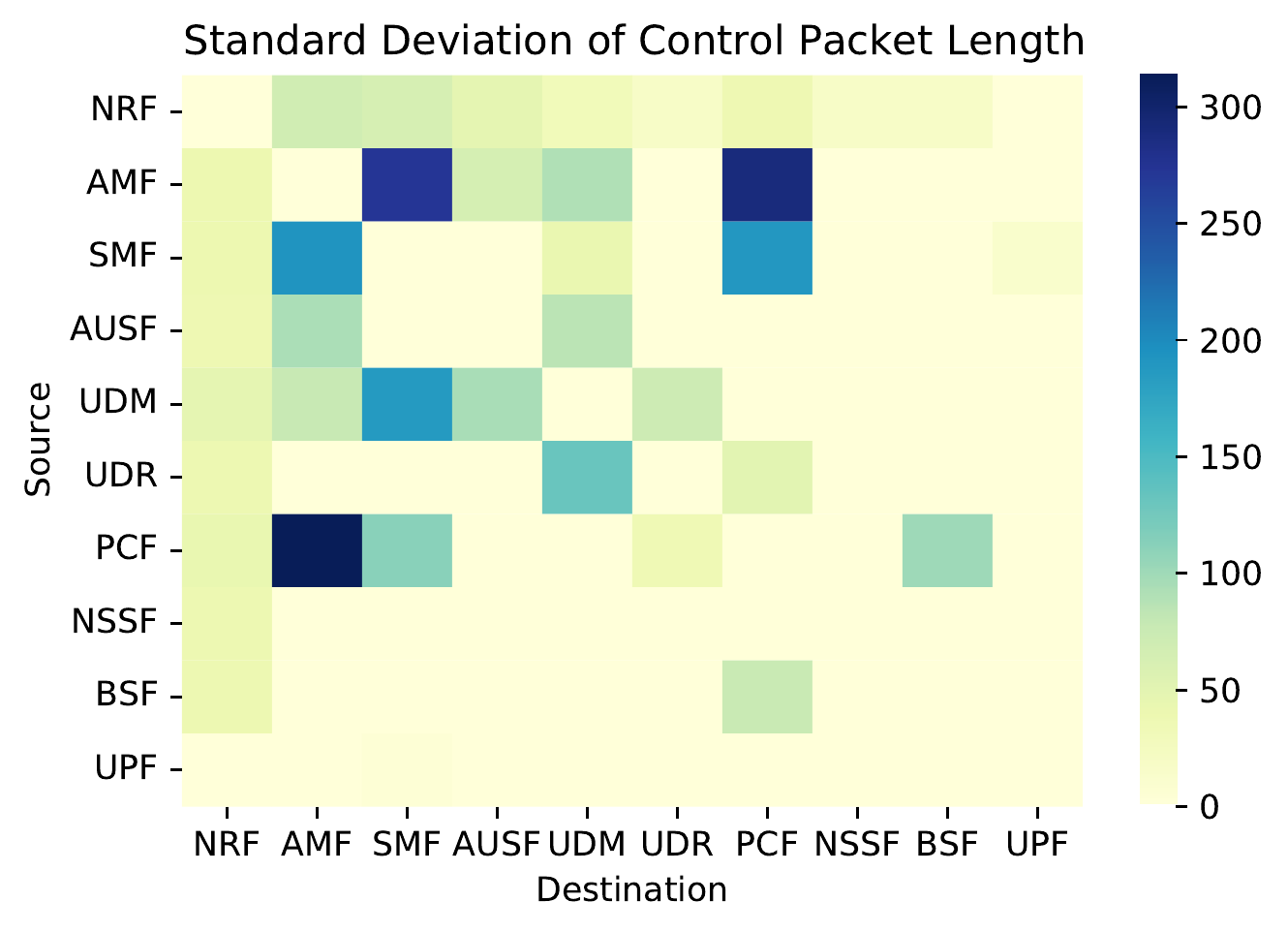}}
\caption{Standard deviation of packet length per NF-NF interaction}
\label{stdev}
\end{figure}

\begin{figure}[!htbp]
\centerline{\includegraphics[width=0.9\columnwidth]{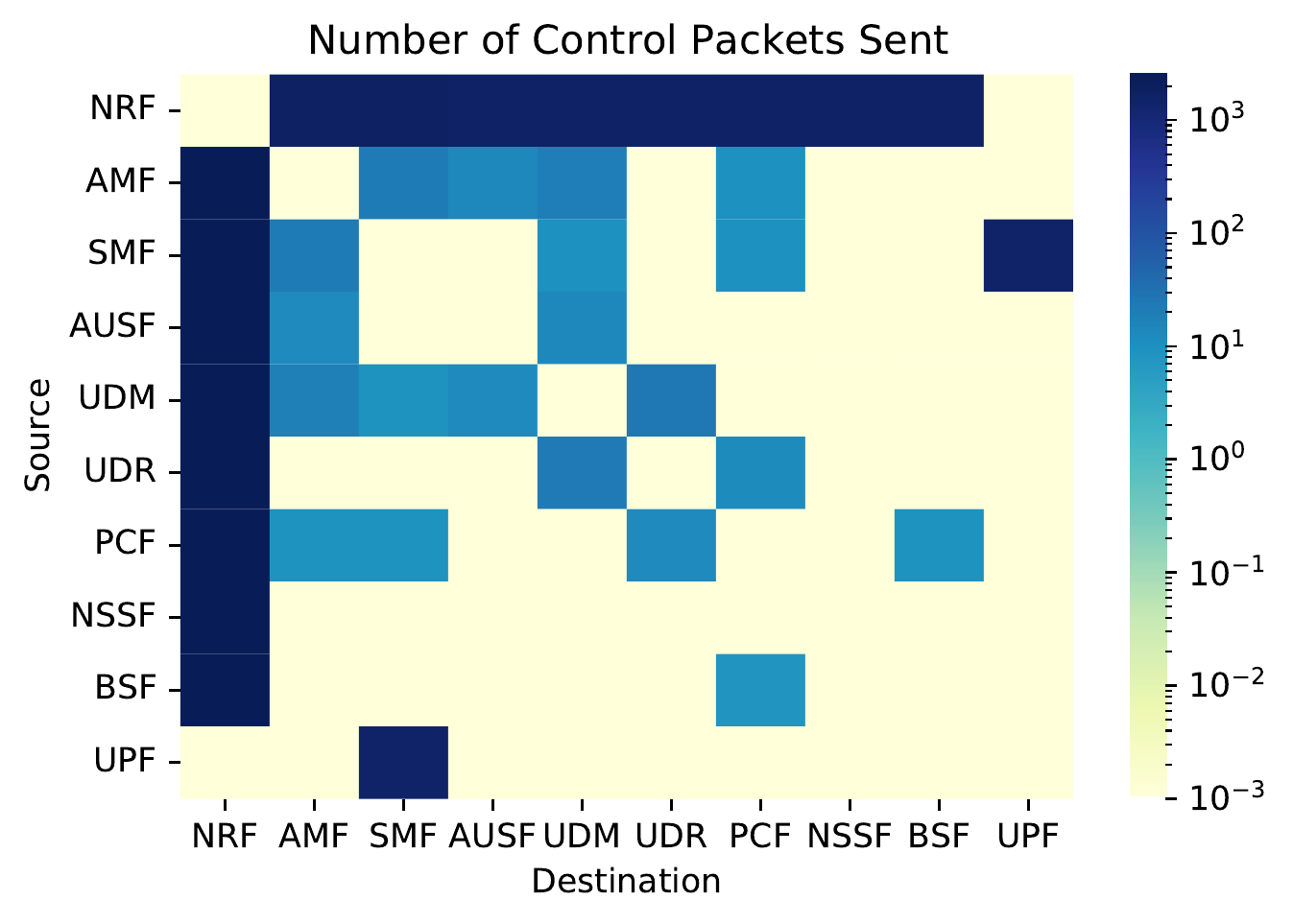}}
\caption{Number of packets sent per NF-NF interaction}
\label{count}
\end{figure}

The various control packet statistics are presented in Fig. \ref{average} - \ref{count}. A heatmap is used to visualize the NF-NF interactions, where each square’s value is expressed through the corresponding colour associated with the colour bar on the right side of each figure. The two-way interactions are expressed through the source and destination NFs denoted by the y and x-axes. \par

As seen in Fig. \ref{average}, the average packet length (bytes) for each NF-NF interaction is displayed. From these results, some key trends and observations emerge. Firstly, the messages sent from the core network functions to the NRF have a higher average packet length than those sent from the NRF to the functions. This observation can be attributed to the initial JSON object sent to the NRF through a PUT request during initial NF registration. This initial registration is part of the NRF NF Management Service, and the received JSON object contains all information related to the NF. The remainder of the interactions are mainly attributed to NF heartbeats to the NRF used to update registration status and state. \par 
Another interesting interaction is that of the PCF and AMF, as there is a slight asymmetry in the average packet length sent. This is attributed to the Access and Mobility Policy Control Service, which creates an individual policy association. When the initial JSON object is sent to the PCF through a POST request, it responds with the same object and added information related to the created policy, thereby creating the asymmetry in the exchanged data. A similar asymmetry is observed with the AMF-PCF interaction. Initially, the AMF uses the SMF PDU Session Service by sending a POST request with a JSON object to create Session Management Context. Once created, the SMF’s response contains limited information, a fraction of what was sent to it, resulting in the observed asymmetry. \par

Figure \ref{max} displays the maximum packet length (bytes) sent across each NF-NF interaction. This statistic was selected as it is an indication of the peak demand of each interaction. As previously discussed, the initial NF registration with the NRF, especially for the functions with more parameters, such as the AMF and SMF, is attributed to the maximum packet length. However, in the case of the AMF, SMF, and AUSF, the NRF sends a larger maximum packet size. This is attributed to the NRF creating subscriptions using its NF Management Service (through a POST request), where it notifies existing functions of newly registered and dependent functions (\textit{i.e.,} AMF is notified of SMF, AUSF, UDM, PCF registration as specified through the \textit{allowedNfTypes} parameter of each of the registered NFs). \par

The next statistic, the standard deviation of the packet length (bytes) exchanged through each NF-NF interaction, is presented in Fig. \ref{stdev}. This figure shows that some NF-NF interactions exhibit consistent information transfer, characterized by low standard deviations, whereas others, such as the AMF/SMF interaction, exhibit higher standard deviations. The standard deviation in the AMF/SMF interaction can be specifically attributed to the combination of packets from the establishment of the initial PDU session and all modification requests, which include additional information exchanges along with the short TCP control messages (\textit{i.e.,} ACK). \par

The final statistic explored is the total number of packets exchanged per NF-NF interaction, as displayed in Fig. \ref{count}. It should be noted that this figure uses a logarithmic scale for the heatmap colour bar. This statistic was selected as it is a good indicator, along with the other presented statistics, of control signalling traffic volume. As expected, the NF-NRF interactions exhibit the greatest number of control packets sent due to the periodic heartbeat used to update NF registration data. Additionally, the SMF/UPF interaction exhibits a high number of packets exchanged. This interaction uses UDP packets and the Packet Forwarding Control Protocol (PFCP), a protocol introduced in 5G Core Networking, to perform Control and User Plane Separation (CUPS). Aside from the initial session establishment and any subsequent session modification requests, this interaction uses periodic heartbeats, similar to the NF-NRF interactions, explaining the high number of packets exchanged. 

\subsection{NRF Interactions}
\begin{figure}[!htbp]
\centerline{\includegraphics[width=0.9\columnwidth]{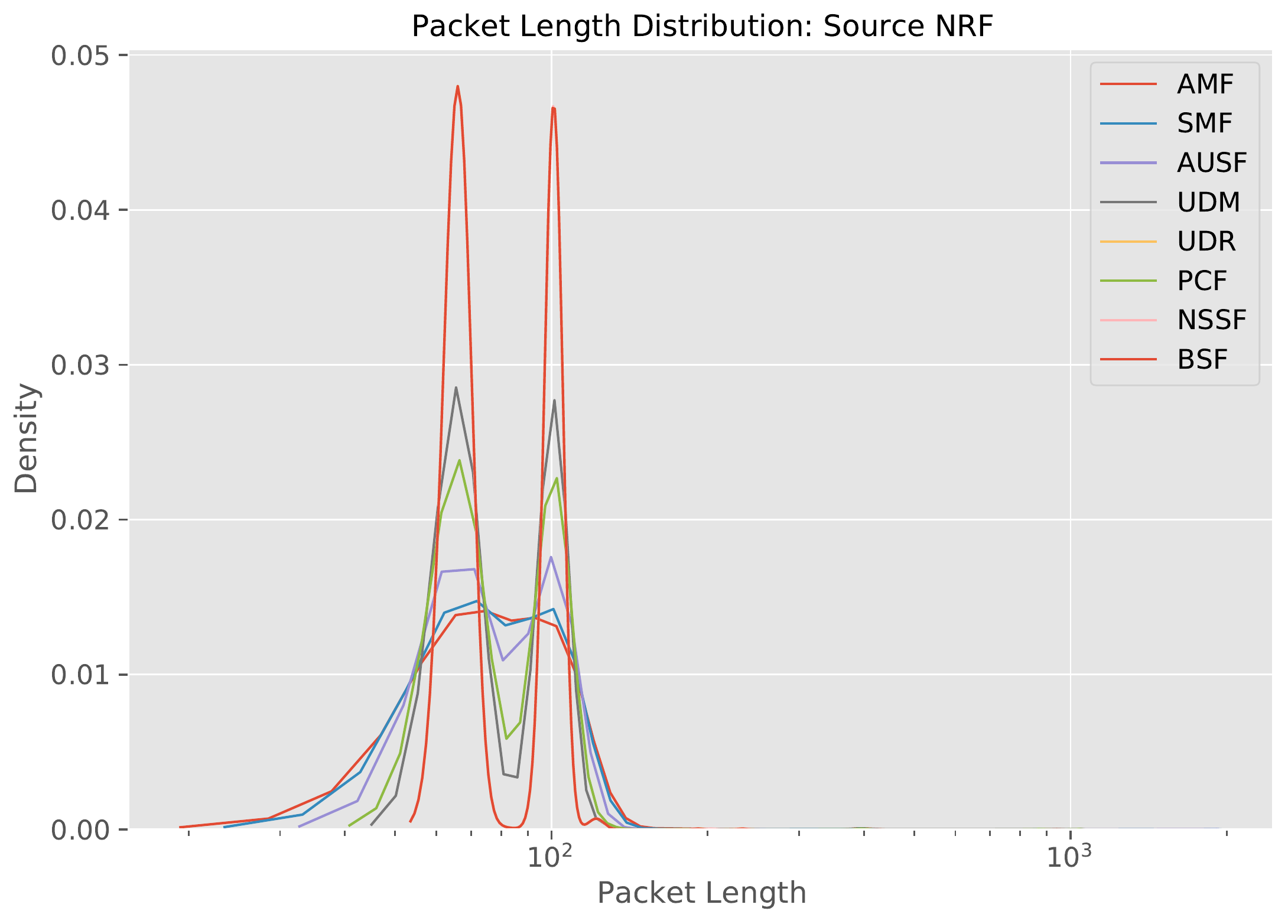}}
\caption{Packet Length Distribution: Source NRF}
\label{source}
\end{figure}

\begin{figure}[!htbp]
\centerline{\includegraphics[width=0.9\columnwidth]{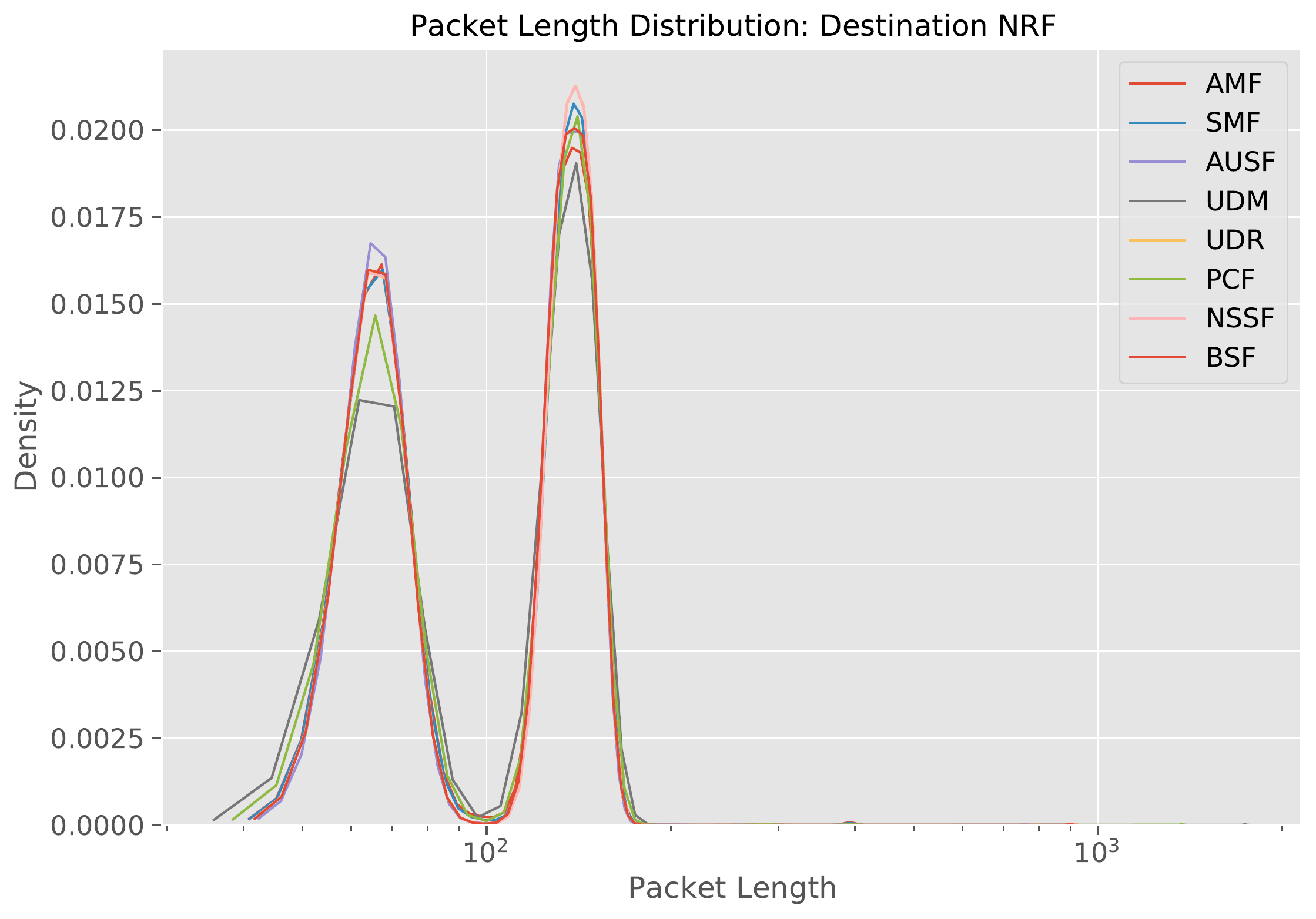}}
\caption{Packet Length Distribution: Destination NRF}
\label{dest}
\end{figure}

Given the results presented in the previous section, the next set of presented results pertains to the interaction of the NRF with the remaining network functions. Two distributions of the length of packets (bytes) sent to and from the NRF are generated to explore this interaction further. Fig. \ref{source} presents the NRF as a source distribution, whereas Fig. \ref{dest} presents the NRF as a destination distribution. \par

As seen in Fig. \ref{source}, for the majority of the network functions, aside from the AMF and SMF, two distinct peaks emerge. With the lower packet length, the left peak corresponds to TCP acknowledgement messages that are sent in response to requests made by the other NFs. With the higher packet length, the right peak corresponds to an HTTP 204 status code response to the NF heartbeat with no content. In the case of the AMF and the SMF, the peaks are less pronounced as there is additional subscription signalling, as previously discussed. \par

Conversely, Fig. \ref{dest} displays the distribution of packet lengths when the NRF is the destination. Aside from the initial NF registration, the remaining interactions from the NFs to the NRF are for heartbeat and status update purposes. As with the previous results, the left peak is associated with TCP acknowledgement packets, whereas the right peak is associated with the NF heartbeat, which is conducted using an HTTP PATCH request with a JSON object containing the current NF registration status as well as any changes in its configuration.

\subsection{Clustering Analysis}
\begin{figure}[!htbp]
\centerline{\includegraphics[width=0.9\columnwidth]{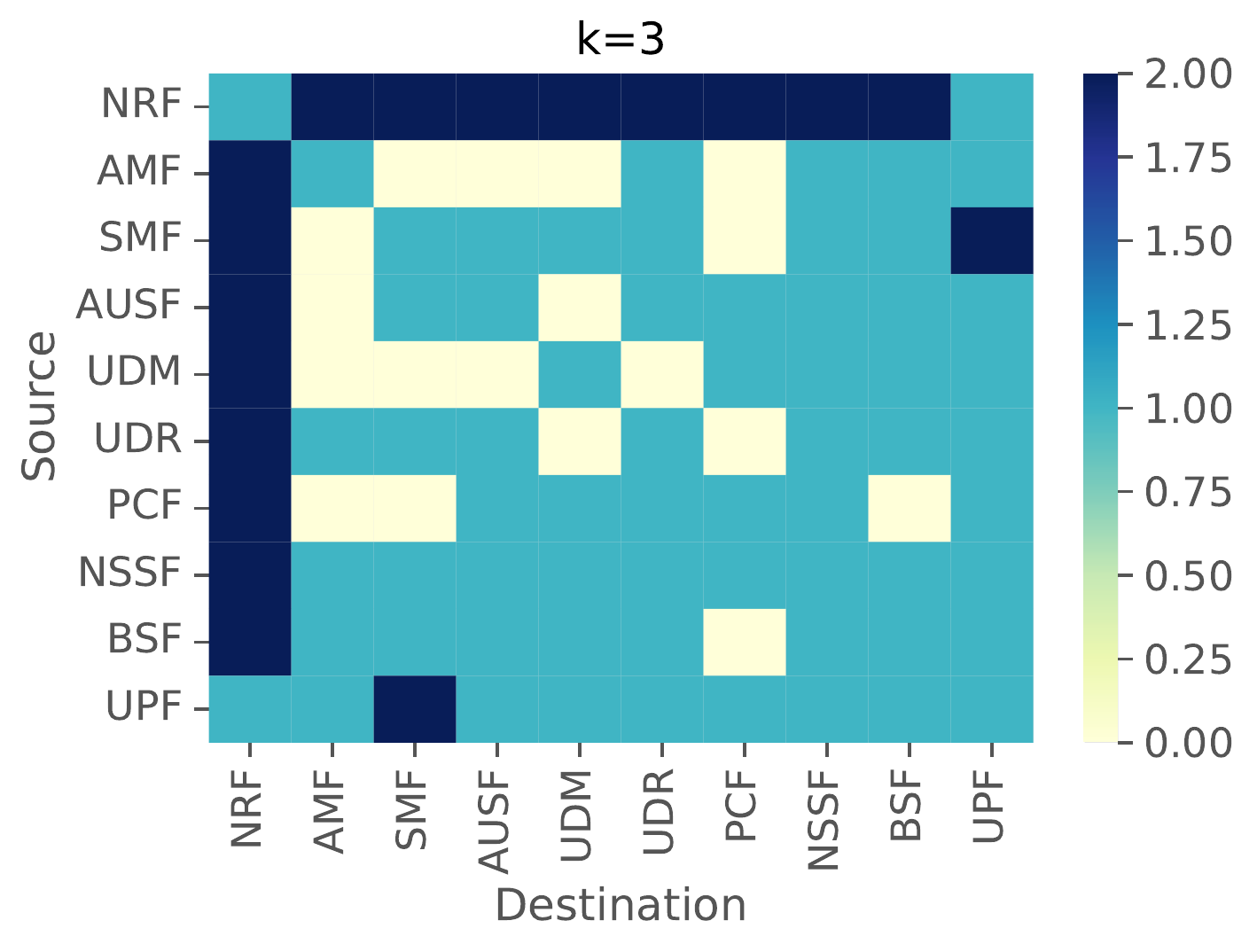}}
\caption{Clustering Analysis: k=3}
\label{k3}
\end{figure}

\begin{figure}[!htbp]
\centerline{\includegraphics[width=0.9\columnwidth]{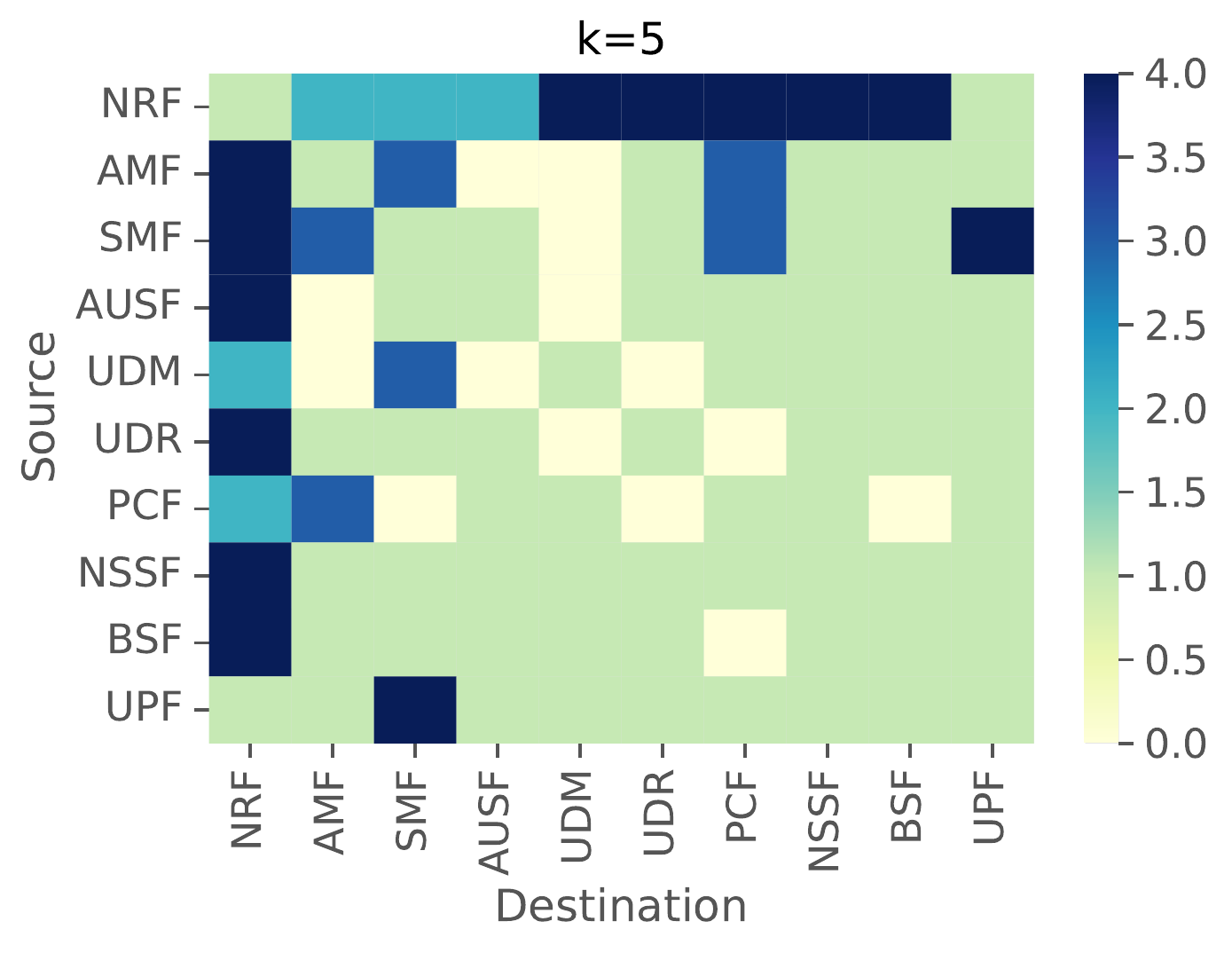}}
\caption{Clustering Analysis: k=5}
\label{k5}
\end{figure}

\begin{figure}[!htbp]
\centerline{\includegraphics[width=0.9\columnwidth]{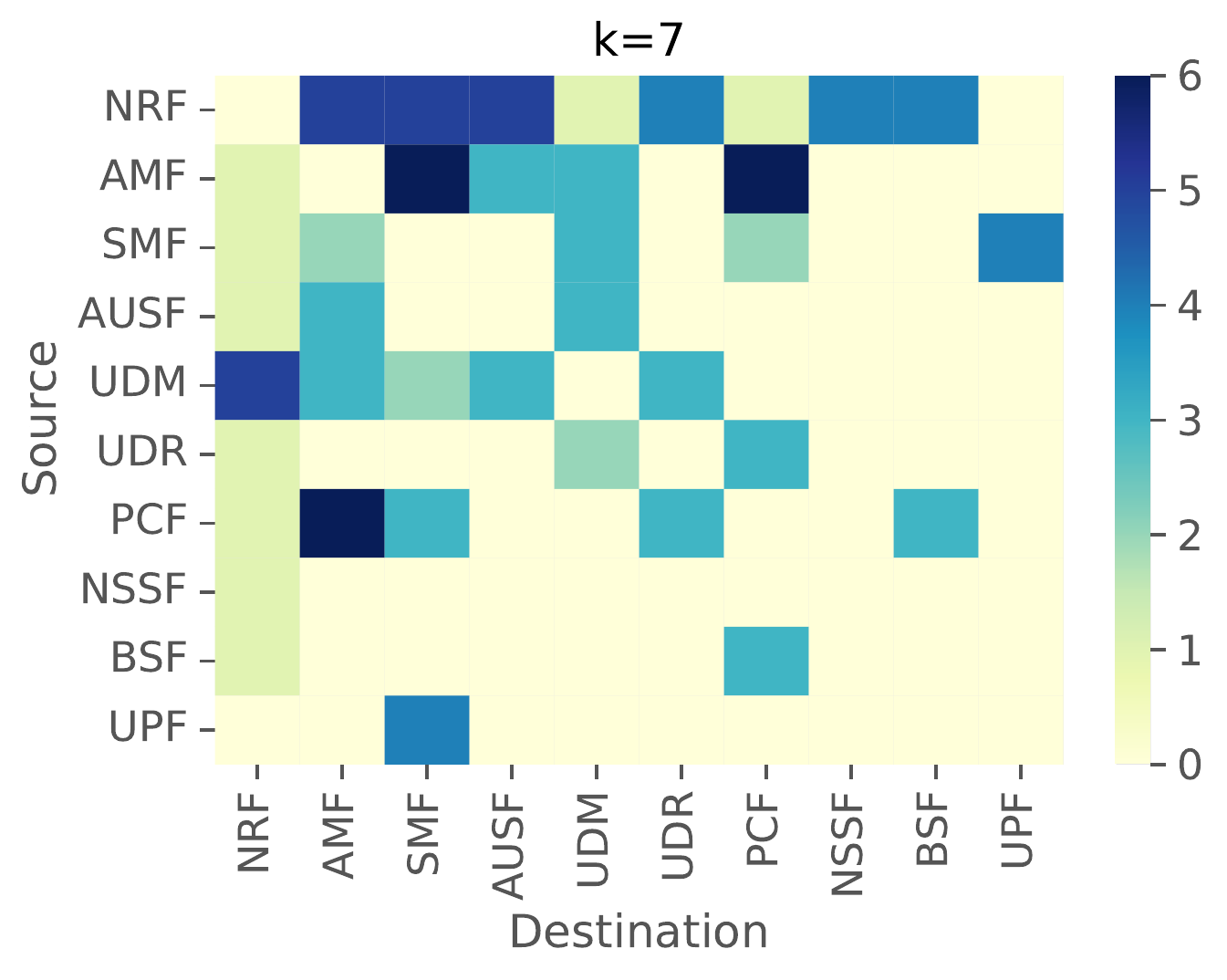}}
\caption{Clustering Analysis: k=7}
\label{k7}
\end{figure}

 The final results presented in this paper pertain to the clustering analysis. As mentioned, the objective of the clustering is to determine similarities between NF-NF interactions, which can be leveraged to make intelligent MANO decisions. For brevity, only some of the clustering results are presented. Figures \ref{k3}, \ref{k5}, and \ref{k7} present the clustering results for $k=3$, $k=5$, and $k=7$, respectively. A heatmap was used to display the results where the colour bar to the right of the map represents the associated cluster label. \par

As seen in Fig. \ref{k3}, the NF-NF interactions are categorized into three clusters with labels 0,1,2. Cluster 2, which includes all NF-NRF interactions and the UPF-SMF interactions, can be characterized by a high number of packets exchanged, as discussed in previous results. Cluster label 1 includes all of the instances where no packets are exchanged and, therefore, no interaction is present; however, it also includes the SMF-UDM interaction. This suggests that $k=3$ is not an ideal number of clusters as this interaction intuitively does not belong in this grouping. Cluster label 0 contains NF-NF interactions that exhibit fewer packets exchanged compared to the NRF-NF interactions. \par

As seen in Fig. \ref{k5}, since $k=5$, the cluster labels now range from 0 to 4. These results show improvement compared to those where $k=3$, as cluster 1 only contains all NF-NF pairs with no packet exchange and, therefore, exhibit no interaction. Furthermore, clusters 0 and 2 from the $k=3$ analysis have each been divided into two resulting clusters. Comparing these results to Fig. \ref{max}, it can be seen that these new cluster divisions are correlated to the maximum packet length exchanged in the interaction. \par

Figure \ref{k7} presents the results of $k=7$ and has cluster labels ranging from 0 to 6. This clustering analysis further divides the $k=5$ clusters, and new patterns emerge. Specifically, in this case, the $k=3$ cluster containing the NF-NRF interactions is now divided into three unique clusters. As with the $k=5$ case, all the NF-NF pairs with no interaction have been successfully grouped. The remaining NF-NF interactions have been divided into three clusters, one notably containing the AMF/PCF interaction and the one-way interaction of the AMF to the SMF. As explained in the following section, the insights gained through the clustering process can be used to prioritize certain interactions during MANO actions, such as VNF scaling and migration. \par


\section{Insights and Opportunities}

\textbf{Intelligent Networking:} The development of the NWDAF is a critical step towards data-driven and intelligent networking. The multitude of insights that can be extracted from the collected data can enhance MANO activities and improve network performance and efficiency. NF scaling and migration are critical MANO functionalities essential for optimizing the network in terms of performance and resource efficiency. To this end, by exploring the interaction between NFs, informed decisions can be made when determining which NFs need scaling and where specific NFs should be migrated to. The objective of such a task could be to perform load balancing by introducing replicate NF instances or migrating existing instances. By characterizing the various NF-NF interactions, it will be possible to determine the optimal placement of the replicate instance to meet the required objective. \par

\textbf{Beyond 5G:} The NWDAF also presents developmental opportunities in the advent of Beyond 5G networks. Beyond 5G networks, similarly to 5G networks, must provide enhanced reliability and low latencies to services and end consumers, both in private and cloud environments. Moreover, Beyond 5G networks are considered to be cloud-native, supporting edge computing in a distributed architecture. Hence, the 3GPP introduced the Management Data Analytics Function (MDAF) to complement NWDAF functionalities by aggregating fault network statistics and performances per cell/region \cite{samdanis2020road}. The system prototype presented in this paper meets a number of conditions that the MDAF mandates for 5G and Beyond networks, including monitoring network function statuses and appending durations to historic data. Compared to the NWDAF, the system prototype creates the opportunity to scale 5G core network resources by anticipating traffic load changes through forecasting via machine learning techniques \cite{alawe2018improving}.\par

\textbf{Resource Allocation:} Furthermore, the NWDAF data analysis has major implications for network resource allocation tasks. By actively monitoring the resource requirements of each NF, the NWDAF can proactively perform resource scaling to prevent an overprovisioned NF from impacting the performance of the network. Additionally, the analysis of specific NF-NF interactions can lead to refined and accurate estimates of network resources requirements (\textit{i.e.,} bandwidth). Also, the clustering of NF-NF interactions can be used as a method of prioritization, where a cluster of resource-intensive interactions will be given allocation priority. \par

\textbf{Network Modelling:} Finally, the collected data can be used to build models of normal (expected) network performance. Once constructed, these models can be used for anomaly detection tasks to identify and mitigate abnormal and adverse network conditions. Additionally, with the prospect of autonomous and intelligent agents, measures need to be taken to ensure that phenomena such as model drift are detected and remediated in a timely fashion. The data collected at various stages of network operation can be used to construct concepts and can be leveraged to perform concept drift detection \cite{manias2021}. 


\section{Conclusion}
This paper has demonstrated key challenges for 5G and Beyond networks towards handling massive amounts of data and demanding requirements. In terms of the NWDAF, it can provide a variety of information and statistics on user data and network-operator data for different NFs. We have detailed the use of a functional NWDAF prototype and fully operational 5GC system prototype, to directly analyze these data interactions and core network signalling traffic in a data-driven approach, provided by the Open5GS and UERANSIM implementations, and Apache Kafka for the data pipelining and event-streaming platform. Machine learning techniques were then applied to the conglomerated data in the fashion of the corresponding NWDAF service. In particular, the analysis of network-generated data to explore core NF-NF interactions was explored, and the use of unsupervised learning to group core NF-NF interactions. Furthermore, the results of the paper demonstrate the use of clustering to determine similarities between NF-NF interactions and in addition, the application of NWDAF insights for Intelligent Networks and NFV MANO. \par

For the NWDAF and 5GC system prototype, future work will continue to explore data generated from 5G core network functions and emulate full service-based functionalities of the NWDAF as specified by the 3GPP. Specifically, more recent services focus on ML model provisioning and information gathering and processing into events and analytics info body responses in the service-based architecture. Finally, forecasting and proactive network management are at the forefront of development of advanced analytics models that will be considered for more use cases using the generated data from the system prototype.

\bibliographystyle{unsrt}
\bibliography{sample}

\end{document}